\documentclass{IEEEtran4PSCC}

\usepackage{gensymb}
\usepackage{comment}
\usepackage{color}
\usepackage{cite}
\usepackage{amsmath, amstext, amssymb,bm}
\usepackage{array}
\usepackage{stfloats}
\usepackage{url}
\usepackage{graphicx}
\usepackage{subfigure}
\newcommand{\squeezeup}{\vspace{-2.5mm}}

\usepackage{tikz}
\usetikzlibrary{%
 arrows,
 positioning
}

\IEEEoverridecommandlockouts

\usepackage{multicol,lipsum}

\begin{document}

\title{Optimal Ensemble Control of Loads in Distribution Grids with Network Constraints} % I would consider changing the title, btw. Forgot to include in my email. What do you think? -Y

\begin{comment}
\author{
\IEEEauthorblockN{Michael Chertkov, Deepjyoti Deka}
\IEEEauthorblockA{Theory Division\\
Los Alamos National Laboratory\\
Los Alamos, NM, USA\\
\{chertkov,deepjyoti\}@lanl.gov }
\and
\IEEEauthorblockN{Yury Dvorkin}
\IEEEauthorblockA{Department of Electrical and Computer Engineering \\
Tandon School of Engineering \\
New York University\\
New York, NY, USA\\
dvorkin@nyu.edu }
}
\end{comment}

\vspace{-0.4cm}
\author{
\IEEEauthorblockN{
Michael Chertkov \IEEEauthorrefmark{1}\IEEEauthorrefmark{2}, Deepjyoti Deka \IEEEauthorrefmark{1} and Yury Dvorkin \IEEEauthorrefmark{3}}
\IEEEauthorblockA{\IEEEauthorrefmark{1} Theory Division,
Los Alamos National Laboratory,\\
Los Alamos, NM, USA,
(deepjyoti, chertkov)@lanl.gov }
\IEEEauthorblockA{\IEEEauthorrefmark{2} Skolkovo Institute of Science and Technology, 143026 Moscow, Russia }
\IEEEauthorblockA{\IEEEauthorrefmark{3} Department of Electrical and Computer Engineering,
Tandon School of Engineering,\\
New York University,
New York, NY, USA, dvorkin@nyu.edu }
}
\vspace{-0.4cm}

\maketitle
\vspace{-1cm}
\begin{abstract}
Flexible loads, e.g. thermostatically controlled loads (TCLs), are technically feasible to participate in demand response (DR) programs. On the other hand, there is a number of challenges that need to be resolved before it can be implemented in practice en masse. First, individual TCLs must be aggregated and operated in sync to scale DR benefits. Second, the uncertainty of TCLs needs to be accounted for. Third, exercising the flexibility of TCLs needs to be coordinated with distribution system operations to avoid unnecessary power losses and compliance with power flow and voltage limits. This paper addresses these challenges.

We propose a network-constrained, open-loop, stochastic optimal control formulation. The first part of this formulation represents ensembles of collocated TCLs modelled by an aggregated Markov Process (MP), where each MP state is associated with a given power consumption or production level. The second part extends MPs to a multi-period distribution power flow optimization. In this optimization, the control of TCL ensembles is regulated by transition probability matrices and physically enabled by local active and reactive power controls at TCL locations. The optimization is solved with a Spatio-Temporal Dual Decomposition (ST-D2) algorithm. The performance of the proposed formulation and algorithm is demonstrated on the IEEE 33-bus distribution model.
\end{abstract}


\thanksto{ The work was supported by funding from the U.S. DOE/OE as part of the DOE Grid Modernization Initiative.}

\section{Introduction}

Demand Response (DR) is an emerging technology that enrolls such flexible loads as Thermostatically Controlled Loads (TCLs) to provide various grid support services. Albeit their uncertainty, TCLs can be controlled without irreversibly compromising their comfort (utility) settings and technical constraints \cite{11CH,14Siano}, while providing frequency control \cite{80STKOPC,10DHMT,13AKCAS,16ZHZM}, ancillary services \cite{DR_ancillary}, congestion management \cite{12YNZM} under different market designs \cite{16ECPRH, 7328775}. The System Operator (SO) can control individual TCLs via sending a target consumption or price signal in a multi-period manner with a given update frequency. The updates can be communicated ahead of time (e.g., for the next 10 min) or in real time (e.g., every 10 seconds), thus enabling fine-grain power delivery. Alternatively, TCLs can be controlled by a third-party aggregator that serves as a mediator between the SO and TCLs \cite{79CD,81IS,84CM,85MC,88MC,04LC,05LCW,09Cal,11CH,11BF}. Introducing the aggregator enables a hierarchical control scheme that improves grid performance and, at the same time, relieves the SO from the prohibitively expensive direct control of TCLs.

In line with the formulation presented below, cycling flexible loads such as TCLs \cite{81IS,84CM,09Cal,11CH,12AK,12MC,15GMK,17CCa} have been modeled using a Markov Decision Process (MDP) in \cite{16BMa,16BMb,17CCb}. The MDP is a discrete-time, discrete-space framework that optimizes a Markov Process (MP) representing stochastic dynamics of TCL ensembles. The MDP scheme is favorable due to its computational and analytic tractability rendered by solution techniques based on dynamic programming. The related previous work in \cite{16BMa,16BMb,17CCb} is prone to the common caveat of neglecting the objective function and constraints of the distribution system operations, e.g. power loss minimization, power flow and voltage limits. As penetration levels and transactivity of TCL ensembles, as well as their geographical diversity, increase, these objective function and constraints might be compromised. For example, exercising the TCL flexibility may cause additional power losses in the power grid, violations of distribution system constraints and, at the same time, lead to suboptimal values of the comfort (utility) function of the TCL ensemble. This paper confronts the caveat of neglecting distribution system operations and proposes an open-loop, stochastic optimal control problem that accounts for the common distribution system objective function and constraints.

\subsection{Contributions}
\begin{figure}[t]
\centering
\includegraphics[scale=0.18,page=3]{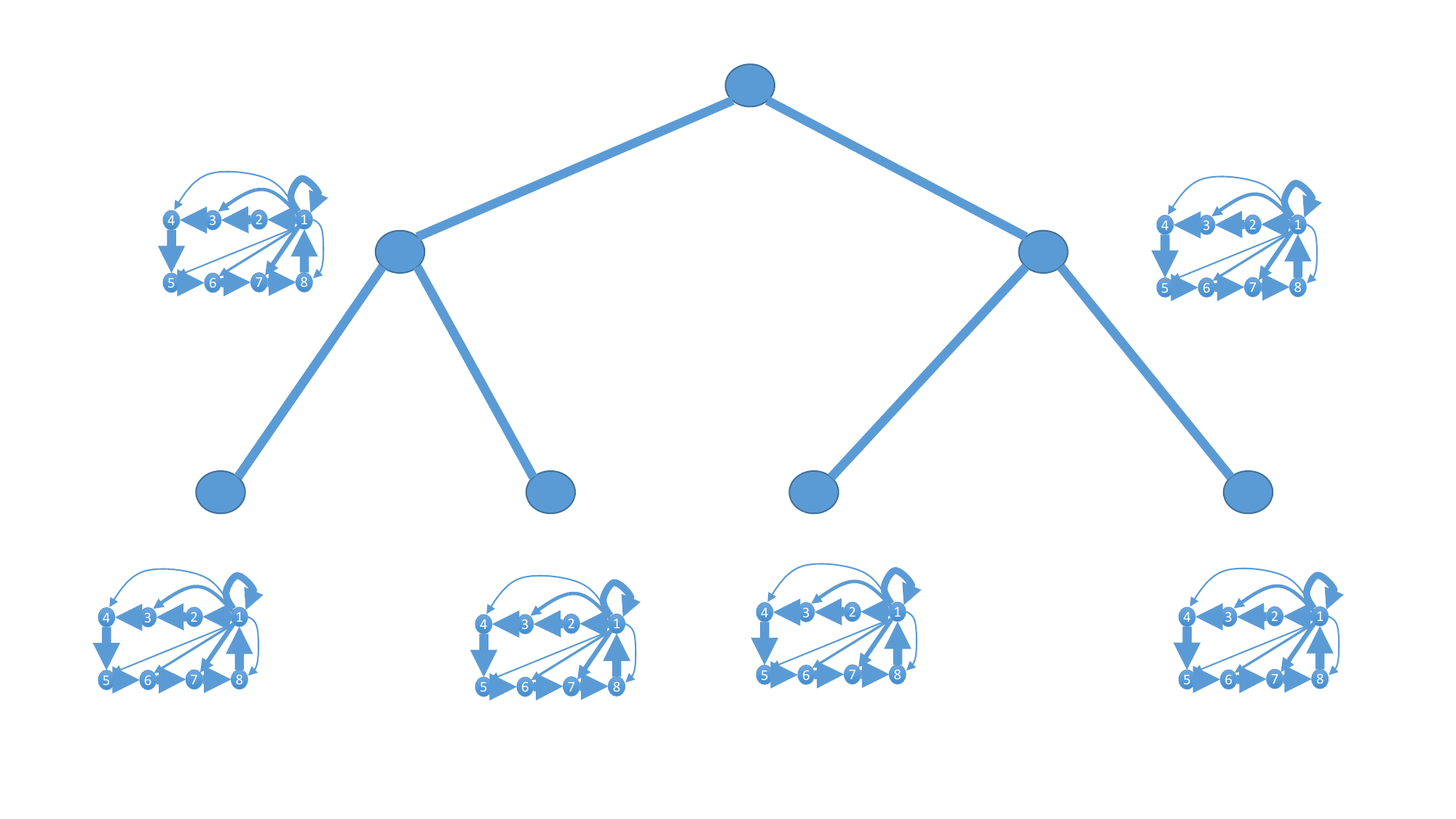}
\caption{Schematic illustration of the distribution system, where some buses feature TCL ensembles that are modeled as a Markov Process. The Markov Process is formally defined in Section~\ref{sec:ensemble} and possible transitions for each ensemble are shown in Fig.~\ref{fig:TCL_MDP}.}
\label{fig:MDP+grid}
\end{figure}

We consider multiple TCL ensembles collocated within a given distribution system, as schematically shown in Fig.~\ref{fig:MDP+grid}. Each ensemble optimizes its own objective function that trade-offs the cost of energy consumed and user-defined comfort (utility) functions of participating TCLs. On the other hand, the system operator aims to optimize its own objective function (e.g. power loss or operating cost minimization), while maintaining operationally feasible line flows and voltage magnitudes. The main contributions of the proposed open-loop, stochastic optimal control formulation are as follows:
%The most significant feature of our open-loop stochastic optimal control formulation, resolved via Spatio Temporal Dual Decomposition (ST-D2) algorithm, is that it is very general. Indeed it can be viewed as an extension of several distinct efforts in distribution grid optimziation.
\begin{itemize}
\item \textcolor{black}{Our formulation enhances  \cite{14SBC, 17CCb, 89BWa} by modeling} temporal and spatial heterogeneity of the TCL ensembles, while solving the MDP and distribution power flow optimization. To accommodate a large number of TCLs in the MDP optimization, we invoke the ``thermodynamic limit'' approximation for large ensembles and design tractable optimal policies for each ensemble.
\item Relative to \cite{15Meyn, 17CCa, 17CCb}, we couple the MDP optimization with the distribution power flow optimization  to explicitly account for ac power flow and voltage constraints and co-optimize the distribution system operations and decisions on TCL ensembles.
\item Our formulation considers time-variable electricity prices and is extensible to account for multiple price forecasts with a receding prediction horizon.
%. Thus this is an open loop control with the spread over time decision computed before the start of the actual execution. %Implementation wise, the open loop dynamic optimization/control is to be implemented via the on-line technique called residing/sliding horizon where the prediction horizon is shifted forward as time progresses.)
\item The proposed formulation is solved using an iterative Spatio-Temporal Dual Decomposition (ST-D2) algorithm. Each iteration is organized as follows. First, the algorithm solves spatially separable Markov Decision Process (MDP) problems for individual TCL ensembles. Second, it uses the MDP solution to solve the distribution power flow optimization. Third, the dual variables related to spatio-temporal decisions are updated based on the solution obtained at the current iteration. The iterative process continues until it converges.
\end{itemize}

\subsection{Paper Organization}
The rest of the paper is organized as follows. Section \ref{sec:system} describes the MDP model of the TCL ensembles and integrates this model into the distribution power flow optimization. Section \ref{sec:algo1} and \ref{sec:hybrid} describe two modifications of the proposed iterative ST-D2 algorithm. We present our numerical results in Section \ref{sec:results} and conclude the paper in Section \ref{sec:conclusion}.

\section{Model}\label{sec:system}

\subsection{Preliminaries} We consider TCL ensembles distributed across different nodes of a given radial distribution system. The system topology is given by graph $\mathcal G=(\mathcal V,\mathcal E)$, where $\mathcal V$ and $\mathcal E$ are the set of buses (nodes) and the set of undirected lines (edges). The voltage magnitudes, active and reactive power injections at bus $i \in \mathcal{V}$ are denoted as $v_{i},p_i,$ and $q_i$, respectively. Element $(ij) \in \mathcal{E}$ denotes the line between buses $i$ and $j$, $i\neq j$, and the active and reactive power flows in that line are denoted as $p_{ij}$ and $q_{ij}$. Each line in $\mathcal{E}$ is characterized by its resistance $r_{ij}$ and reactance $x_{ij}$. The root bus of the distribution system is chosen to be the slack bus, where the system maintains the reference voltage and compensates for the power mismatch.

\subsection{TCL Ensembles} \label{sec:ensemble} From the practical point of view, TCL ensembles at different buses of distribution system are likely to be operated \textcolor{black}{by different and independent entities (e.g.  building managers or aggregators)}; these entities are likely to act independently with little, if any, temporal and spatial coupling\textcolor{black}{\footnote{\textcolor{black}{Such correlations may however exist between different TCL assembles and can be accounted within the proposed framework, see \cite{6832599}.} }.} Thus, each TCL ensemble located at a bus of the distribution system can be modeled using a discrete-time, discrete-space MP similar to \cite{07Tod,12DjEmo,13DjEmo,17CCb}. In the following, we consider a finite time horizon of operation, i.e. $0\leq t \leq T$. The dynamic state of each TCL in the ensemble at bus $i$ is characterized by vector $\rho_{i}(t)=(\rho^\alpha_{i}(t)|\forall \alpha)$, where $\rho^\alpha_i(t)\geq 0$ is the probability vector that indicates that TCLs at bus $i$ are in state $\alpha$. The probability vector is normalized over a given time horizon as $\sum_{\alpha} \rho^\alpha_i(t)=1,\quad\forall t{=0 ,\cdots, T},~~\forall \alpha$. We define the MP at each bus $i$ using the transition probability matrix $\mathcal{P}_i(t)=(\mathcal{P}^{\alpha\beta}_i(t)|\forall t,\forall \alpha,\beta)$, where $\mathcal{P}^{\alpha\beta}_i(t)$ is the vector that characterize the probability of the transition of TCLs at node $i$ from state $\beta$ at time $t$ to state $\alpha$ at time $t+1$. We also impose the integrality/stochasticity constraint on possible transitions as:
\begin{align}
\sum_\alpha \mathcal{P}^{\alpha\beta}_i(t)=1,\quad \forall t=0,\cdots, T-1,~~ \forall \beta,
\label{stochastic}
\end{align}
The temporal evolution of the TCL ensemble at node $i$ can then be modeled as:
\begin{align}
\rho^\alpha_i(t+1)=\sum_\beta \mathcal{P}^{\alpha\beta}_i(t)\rho^\beta_i(t),\quad \forall t,~~ \forall \alpha,
\label{master-eq}
\end{align}
where the initial condition ($\rho^\alpha_{in;i}$) is enforced as:
\begin{eqnarray}
\rho_i^\alpha(0)=\rho^\alpha_{in;i},\quad \forall \alpha.
\label{rho_init}
\end{eqnarray}
Using \eqref{stochastic}-\eqref{rho_init}, the MDP optimization for the ensemble at bus $i$ can be stated as:
\begin{align}
\footnotesize
&\min_{\mathcal{P}_i,\rho_i} \mathbb{E}_{\rho_i}\sum_{t=0}^{T-1} \sum_\alpha %U^\alpha_i(t+1)+\sum_\beta {\gamma^{\alpha\beta}}_i(t)\log \frac{\mathcal{P}^{\alpha\beta}_i(t)}{\bar{\mathcal{P}^{\alpha\beta}}_i}\\
\underbrace{U^\alpha_i(t+1)}_{\mbox{Cost of energy}}+ \nonumber \\ & \hspace{4cm}
\underbrace{\sum_\beta { \gamma_i^{\alpha\beta}}(t)\log \frac{\mathcal{P}^{\alpha\beta}_i(t)}{\overline{\mathcal{P}}^{\alpha\beta}_i}}_{\mbox{Welfare penalty}} \label{mdp_obj}\\
&\text{s.t.~} \text{Eq.~(\ref{stochastic},\ref{master-eq},\ref{rho_init})}\label{profit_vs_welfare}
\end{align}
where the matrix $\mathcal{P}^{\alpha\beta}_i (t)$ is the matrix of decision variables that optimizes the state of individual TCLs in the ensemble and $\overline{\mathcal{P}}$ is an exogenous matrix describing the transition probabilities corresponding to ``normal'' (e.g. user-defined) TCL dynamics within the ensemble. Eq.~\eqref{mdp_obj} aims to trade-off the expected cost of energy consumed ($U_i^{\alpha}$) by the TCL ensemble at node $i$ and the welfare penalty. The welfare penalty represents the discomfort (utility loss) caused by the difference between the optimized ($\mathcal{P}^{\alpha\beta}_i$) and ``normal'' ($\overline{\mathcal{P}}^{\alpha\beta}_i$) transition probabilities and is computed based on the Kullback-Leibler (KL) distance \cite{coverthomas,07Tod}. KL distance is one of the key metrics used for computing deviations between two probability distributions in different statistical and engineering fields. For identical distributions, the KL distance is zero and it is non-negative otherwise. Note that weighting parameters $\gamma^{\alpha\beta}_i(t)$ differentiate between the transitions between any two states at each time step and can selectively influence the KL distance. As discussed in Appendix~\ref{app:backforward}, our choice of the KL distance also aids to solve the MDP optimization in \eqref{mdp_obj}-\eqref{profit_vs_welfare} using the backward-forward algorithm.

The MDP optimization formulated in \eqref{mdp_obj}-\eqref{profit_vs_welfare} generalizes the family of the so-called linearly solvable (LS) MDPs introduced in \cite{07Tod} and discussed in \cite{12DjEmo,13DjEmo,15Meyn}. We refer interested readers to \cite{07Tod} for further detail.

%The dynamic programming based solution scheme for Problem \ref{profit_vs_welfare} is discussed below.

\subsection{Distribution Power Flow Optimization with MDP}
The MDP optimization formulated in \eqref{mdp_obj}-\eqref{profit_vs_welfare} can be integrated in the distribution system power flow optimization. This optimization aims to minimize active power losses (and hence the operating cost) and considers ac power flows over the radial topology. Unlike the spatially-separable MDP optimization in \eqref{mdp_obj}-\eqref{profit_vs_welfare}, the distribution system power flow optimization is only temporally separable.

The integrated MDP and distribution power flow optimization is then given by:
\begin{eqnarray}
&& \hspace{-1cm}\min\limits_{\begin{array}{c}\rho, {\mathcal P}, v,\\ p,q,\rho,\phi,\\ p_c,q_c\end{array}} \!\sum\limits_{t=0}^{T-1}\!\Biggl[\!
\mu_t \sum_{(ij) \in {\mathcal E}}r_{ij}\frac{p^2_{ij}(t)+q^2_{ij}(t)}{v^2_i(t)}+\nonumber\\&&\hspace{-.7cm}{\sum\limits_i}\!\sum\limits_{\alpha,\beta}\! \begin{footnotesize}{\mathcal P}^{\alpha\beta}_i(t)\left(U^\alpha_i(t+1)\!+ \gamma_i^{\alpha\beta}(t)\log \frac{\mathcal{P}^{\alpha\beta}_i(t)}{{\bar{\mathcal{P}}}^{\alpha\beta}_i}\right)\rho^\beta_{i}(t)\end{footnotesize}\Biggr]
\label{hybrid_1}\\
&&\hspace{-0.6cm}\mbox{s.t. }\rho^\alpha_i(t+1)=\sum_\beta {\mathcal P}^{\alpha\beta}_{i}(t)\rho^\beta_i(t),\ \forall t,\ \forall i,\quad \forall \alpha
\label{hybrid-master_1}\\
&& \sum_\alpha\! p_i^\alpha \rho^\alpha_i(t) = p_i(t),\ \forall t,\ \forall i,
\label{hybrid-PF_1p}\\
&& \sum_\alpha\! q_i^\alpha \rho^\alpha_i(t) = q_{i}(t),\ \forall t,\ \forall i,
\label{hybrid-PF_1q}\\
&&p_{ij}(t)-r_{ij}\frac{p^2_{ij}(t)+q^2_{ij}(t)}{v^2_i(t)}=\nonumber\\ &&
p_i(t)+p^c_i(t)+\sum_{k:(jk) \in \mathcal{E}} p_{jk}(t),\ \forall t,\ \forall i,\label{DFp}\\
&&q_{ij}(t)-x_{ij}\frac{p^2_{ij}(t)+q^2_{ij}(t)}{v^2_i(t)}=\nonumber\\ &&
q_i(t)+q^c_i(t)+\sum_{k:(jk) \in \mathcal{E}} q_{jk}(t),\ \forall t,\ \forall i,\label{DFq}
\end{eqnarray}
\begin{eqnarray}
&&v^2_j(t)=v^2_i(t)- 2 (r_{ij} p_{ij}(t)+x_{ij} q_{ij}(t))\nonumber\\ && -(x_{ij}^2+r_{ij}^2)\frac{p^2_{ij}(t)+q^2_{ij}(t)}{v^2_i(t)},\ \forall t,\ \forall i,\label{DFv}\\
&&\underline{v}_i\leq v_i(t)\leq \overline{v}_i,\ \forall t,\ \forall i.
\label{hybrid-voltage_1}\\
&& \underline{p}_i^c\leq p^c_i(t)\leq \overline{p}_i^c,\ \forall t,\ \forall i,\label{p_c}\\
&& \underline{q}_i^c\leq q_i^c(t)\leq \overline{q}_i^c,\ \forall t,\ \forall i.\label{q_c}
\end{eqnarray}
The first term in \eqref{hybrid_1} represents the power loss minimization objective function as customarily used for distribution systems \cite{89BWa,89BWb,89BWc} and the second term represents the MDP objective function as given in \eqref{mdp_obj}. Note that parameter $\mu_t$ in \eqref{hybrid_1} monetizes the power losses to make them comparable to the MDP objective function. Eq.~\eqref{hybrid-master_1} is identical to \eqref{master-eq} in the MDP optimization. The active and reactive power injections of the TCL ensemble at bus $i$ are computed in \eqref{hybrid-PF_1p} and \eqref{hybrid-PF_1q} based on the rated active ($p_i^\alpha$) and reactive ($q_i^\alpha$) power at state $\alpha$ and its probability $\rho^\alpha_i(t)$. Eq.~\eqref{DFp}-\eqref{DFv} represent the \textit{DistFlow} formulation for ac power flows in radial distribution systems based on \cite{89BWa,89BWb,89BWc}. Eq.~\eqref{hybrid-voltage_1} limits voltage magnitudes between their minimum ($\underline{v}_i$) and maximum ($\overline{v}_i$)
limits. Eq.~\eqref{p_c}-\eqref{q_c} limit decision variables on the nodal active and reactive power injections $p^c_i$ and $q^c_i$ between their  minimum ($\underline{p}^c_i$ and $\underline{q}^c_i$) and maximum ($\overline{p}^c_i$ and $\overline{q}^c_i$) limits.

For the sake of simplicity and without loss of generality with respect to the intended contributions, we approximate the \textit{DistFlow} Eq.~(\ref{DFp}-\ref{DFv}) using the \textit{LinDistFlow} model as described in \cite{89BWa}. The \textit{LinDistFlow} model renders linear expressions instead of second-order terms in Eq.~(\ref{DFp}-\ref{DFv})\textcolor{black}{, at the expense of neglecting power losses in distribution lines}. Even with this simplification, the optimization problem in \eqref{hybrid_1}-\eqref{q_c} can hardly be solved using off-the-shelf algorithmic solution, especially for large TCL ensembles and distribution systems. Therefore, Section~\ref{sec:algo1} describes an iterative solution technique that can solve this optimization problem efficiently and in a decentralized manner.
\vspace{10pt}
\section{ST-D2 Algorithm}
\label{sec:algo1}
We solve the optimization problem in \eqref{hybrid_1}-\eqref{q_c} using the proposed Spatio-Temporal Dual Decomposition (ST-D2) algorithm. The algorithm exploits the separation between spatial and temporal variables to iteratively seek the optimal solution. \textcolor{black}{This separation also makes it possible to simultaneously account for the perspective of TCL aggregators, which are likely to operate TCL ensembles  and thus solve the MDP optimization, and the perspective of the distribution system operator, which needs to account for the impact of TCLs in their power flow optimization. From the implementation perspective, this decomposition  is also useful as it allows to combine the dynamic-programming-based MDP optimization and the power flow optimization within one algorithm.}

To explain the algorithm, we generalize the problem in \eqref{hybrid_1}-\eqref{q_c} in the following form:
\begin{eqnarray}
\hspace{-0.2cm}\min\limits_{x,y}\! \left(\!\sum_i\! A_i(x_i)\!+\!\sum_t\! B_t(y(t))\!\right)_{\!\forall t, i: x_i(t)=C^i_t(y_i(t))},
\label{master-schematic}
\end{eqnarray}
where $x=\big\{\rho, \mathcal{P} \forall i \big\}$ denote the MDP decisions as in \eqref{mdp_obj}-\eqref{profit_vs_welfare} and $y=\big\{v_i,p_{ij},q_{ij},p^c_i,q^c_i ~\forall i,j\big\}$ denote the decisions of the distribution system power flow optimization. The spatio-temporal components are denoted as $x_i(t)$ and $y_i(t)$. Similar to \eqref{hybrid_1}, the first term in \eqref{master-schematic} is spatially separable for each MDP, while the second term in \eqref{master-schematic} is temporally separable into variables entering network optimization at each time instant. The condition ${\forall t, i: x_i(t)=C^i_t(y_i(t))}$ in \eqref{master-schematic} represents (\ref{hybrid-PF_1p}-\ref{hybrid-PF_1q}) that relate temporally and spatially separable variables. Other constraints of \eqref{hybrid_1}-\eqref{q_c} are omitted for the clarity of our explanation. The Lagrangian function of \eqref{master-schematic} is then as follows:
%{\begin{footnotesize}
\begin{align}
&\max_\lambda \min_{x,y} \sum_i A_i(x_i)+\sum_t B_t(y(t))+ \nonumber\\ &\hspace{3.7cm} \sum_{i,t} \lambda^i_t\left(x_i(t)-C^{i}_t(y_i(t))\right), \label{master-schematic-Lagr}
\end{align}%\end{footnotesize}}
where $\lambda$ denote the vector of Lagrange multipliers. Note that \eqref{master-schematic-Lagr} can be augmented by adding the second-order term $\sim \sum_{i,t} \left(x_i(t)-C^{i}_t(y_i(t))\right)^2$ as customarily done in Alternating Direction Method of Multipliers (ADDM) algorithms. However, we do not use the augmentation in this paper as we aim to a linear MDP optimization and use methods for the linearly solvable class.

After initializing $\lambda$ to zero, Eq.~(\ref{master-schematic-Lagr}) can be solved iteratively as explained below:
\begin{itemize}
\item[(1)] Solve the spatially-separable MDP problem for each TCL ensemble:
\begin{eqnarray}
\forall i:\ \min_{x_i}\left(A_i(x_i)+\sum_t \lambda^{i}_t x_i(t)\right),\label{first_step}
\end{eqnarray}
\item[(2)] Solve the temporally-separable distribution power flow optimization problem with MDP:
\begin{eqnarray}
\forall t: \min_{y(t)}\left(B_t(y(t))-\sum_i \lambda^{i}_t C^{i}_t(y_i(t))\right),\label{second_step}
\end{eqnarray}
\item[(3)] Update the Lagrangian multipliers (gradient ascent in dual variables)
\begin{eqnarray}
\hspace{-0.6cm}\forall t,i:\ \lambda^{i}(t)\!\leftarrow\!\lambda^{i}(t)\!+\!\delta^{i}(t)\!\left(\!x_{i}(t)\!-\!C^{i}_t(y_{i}(t))\!\right)\!,
\label{third_step}
\end{eqnarray}
where the gradient coefficients, $\delta^{(i)}(t)$, can be chosen to be constant or spatio-temporally heterogeneous, or adjusted in the interest of getting a better convergence.
\end{itemize}

In terms of the optimization problem stated in \eqref{hybrid_1}-\eqref{q_c}, the ST-D2 algorithm is implemented as follows:
\begin{itemize}
% ... in three steps, but coupled problems, then iterating between the two till convergence with an additional step updating Lagrangian multipliers associated with Eqs.~(\ref{hybrid-PF_1p},\ref{hybrid-PF_1q})). The first sub-problem will consist in
\item [(1)] %Fixing (to current values) $p$ and $q$ and then {\bf solving individual MDPs} (each separately)
{ Solve the MDP for each TCL ensemble}:
\begin{align}
&
\forall i: \min\limits_{\rho, {\mathcal P}}\!\sum\limits_{t=0}^{T-1}\!
\sum\limits_\alpha\!
 {\mathcal P}^{\alpha\beta}_i(t)\bigg(\tilde{U}^\alpha_i(t+1)+ \nonumber \\ & \hspace{2cm} \gamma_i^{\alpha\beta}(t)\log \frac{\mathcal{P}^{\alpha\beta}_i(t)}{{\bar{\mathcal{P}}}^{\alpha\beta}_i}\bigg) \rho^\beta_{i}(t), \label{hybrid_21a} \\
%\left(\!\tilde{U}^\alpha_i(t+1)\!+\!\sum\limits_\beta \log \frac{{\mathcal P}^{\alpha\beta}_i(t)}{\bar{{\mathcal P}}^{\alpha\beta}_i}\!\right)\!\rho^\alpha_i(t),\label{hybrid_2a}\\
& \mbox{s.t.}~~% \\ &
\mbox{Eq.~(\ref{hybrid-master_1})},\label{hybrid_2a}\\
&  \hspace{.7cm}\tilde{U}^\alpha_{i}(t)=U^\alpha_{i}(t)+\lambda^{i}_p(t)p_i^\alpha+\lambda^{i}_q(t)q_i^\alpha,\ \forall t,\ \forall i,
\label{tildeU}
\end{align}
where $\lambda^{i}_p(t)$ and $\lambda^{i}_q(t)$ are the Lagrangian multipliers. Note that at the first iteration $\lambda^{i}_p(t)= \lambda^{i}_q(t) =0$.
\item [(2)] Solve the distribution power flow optimization problem with all MDPs:
\begin{eqnarray}
&& \hspace{-1cm}\forall t:~% \nonumber \\ && \hspace{-1cm}
\min\limits_{\begin{array}{c}v_i,p_i,q_i,p_{ij},q_{ij},\\ p^c_i,q^c_i \forall i,j\end{array}} \sum_{(ij) \in {\cal E}}r_{ij}\frac{p^2_{ij}(t)+q^2_{ij}(t)}{v^2_i(t)}-\nonumber\\
&& \hspace{+2.cm}\sum_i \left(\lambda^{i}_p(t)p_i(t)+ \lambda^{i}_q(t)q_{i}(t)\right)
\label{hybrid_2b}\\
&&\mbox{s.t.} ~~\mbox{Eq.~(\ref{DFp})-(\ref{q_c})},
\nonumber
\end{eqnarray}
\item [(3)] {Update the Lagrange multipliers} according to the dual gradient ascent scheme:
\begin{eqnarray}
&& \hspace{-1.8cm}\lambda^{i}_p(t)\!\leftarrow\! \lambda^{i}_p(t)\!+\!\delta^{i}(t) \!\!\left(\!\!\sum_\alpha\! q_{i}^\alpha\rho^\alpha_{i}(t) \!-\! q_{i}(t)\!\right)\!, \forall t,\forall i,
\label{lambda_p_update}\\
&& \hspace{-1.8cm}\lambda^{i}_q(t)\!\leftarrow\! \lambda^{i}_q(t)\!+\! \delta^{i}(t) \!\!\left(\!\!\sum_\alpha\! q_{i}^\alpha \rho^\alpha_{i}(t) \!-\! q_{i}(t)\!\right)\!, \forall t,\forall i.
\label{lambda_q_update}
\end{eqnarray}
where $\delta$ is an exogenous parameter.
\end{itemize}

The optimization problem (\ref{hybrid_2a})-(\ref{tildeU}) is solved using dynamic programming in a deterministic fashion since individual MDPs are static, in the asymptotic limit, with an \textit{infinite} number of TCLs. \textcolor{black}{In other words, we assume a relatively large portfolio of TCLs at each node, which is expected to materialize in distribution systems of the future. Under this assumption, injections associated with individual MDPs at each state do not fluctuate due to self-averaging. Thus, (\ref{hybrid_2a})-(\ref{tildeU}) can be solved using a backward-forward algorithm described in Appendix \ref{app:backforward}. Currently, penetration levels of TCLs are  modest and therefore the number of TCLs  is \emph{finite}, which implies that MDPs can fluctuate.}  We refer interested readers to Appendix \ref{app:large_n} for details of the asymptotic statistics (law of large numbers) for large and finite MDPs as it is of importance for the future work with uncertainty-aware versions of Eq.~(\ref{hybrid_1}). \textcolor{black}{Modeling a finite number of TCLs is left for our future work.}

% An updated version of ST-D2 algorithm that combines steps (2) and (3) is discussed in the next section.

\section{Hybrid ST-D2 Algorithm}
\label{sec:hybrid}
In this section we describe two modifications of the algorithm in Section \ref{sec:algo1}. First, the MDP decisions obtained in Step 1 are  used in the distribution power flow optimization solved at Step 2. Second, the Lagrangian multipliers can now be updated within Step 2. The modified algorithm, abbreviated in the rest of this paper as ST-Hybrid, is then given by:

\begin{itemize}
\item [(1)] {Solve the MDP for each TCL ensemble:}
\begin{eqnarray}
&& \hspace{-1.4cm}\forall i: \min\limits_{\rho, {\mathcal P}}\!\sum\limits_{t=0}^{T-1}\!
\sum\limits_\alpha\!
 {\mathcal P}^{\alpha\beta}_i(t)\bigg(\tilde{U}^\alpha_i(t+1) + \\ && \hspace{2cm} \gamma_i^{\alpha\beta}(t)\log \frac{\mathcal{P}^{\alpha\beta}_i(t)}{{\bar{\mathcal{P}}}^{\alpha\beta}_i}\big)\rho^\beta_{i}(t)\nonumber\\
&&\hspace{-1.4cm} \mbox{s.t.}~~ %\nonumber \\
%&&\hspace{-1.4cm}
\mbox{Eq.~(\ref{hybrid-master_1})},\label{hybrid_3a}\\
&& \hspace{-.8cm}\tilde{U}^\alpha_{i}(t)=U^\alpha_{i}(t)+\lambda^{i}_p(t)p_i^\alpha+\lambda^{i}_q(t)q_i^\alpha,\ \forall t,\ \forall i.
\end{eqnarray}
\item [(2)] Solve the distribution power flow optimization problem with all MDPs and update the Lagrange multipliers:
\begin{eqnarray}
&& \hspace{-1cm}\forall t:%\nonumber \\ && \hspace{-1cm}
\min\limits_{\begin{array}{c}v_i,p_i,q_i,p_{ij},q_{ij},\\ p^c_i,q^c_i \forall i,j\end{array}} \sum_{(ij) \in {\cal E}}r_{ij}\frac{p^2_{ij}(t)+q^2_{ij}(t)}{v^2_i(t)}
\label{hybrid_3b}\\
&& \hspace{-1cm} \mbox{s.t.}~~% \\ && \hspace{-1cm}
\mbox{Eq.~(\ref{hybrid-PF_1p},\ref{hybrid-PF_1q},\ref{DFp},\ref{DFq},\ref{DFv},\ref{hybrid-voltage_1},\ref{p_c},\ref{q_c})},
\nonumber
\end{eqnarray}
where the right-hand side in constraints (\ref{hybrid-PF_1p},\ref{hybrid-PF_1q}) are given by Step (1). Update $\lambda^{i}_p(t),\lambda^{i}_p(t)$ with values of Lagrange variables for constraints (\ref{hybrid-PF_1p},\ref{hybrid-PF_1q}).
\end{itemize}

%Our experiments with the optimization formulation on an IEEE test model are discussed next.
\vspace{10pt}
\section{Results}
\label{sec:results}
\subsection{Data \& Experimental Setup} \label{sec:data} 
\begin{figure}[b]
\centering
\includegraphics[width=0.40\textwidth]{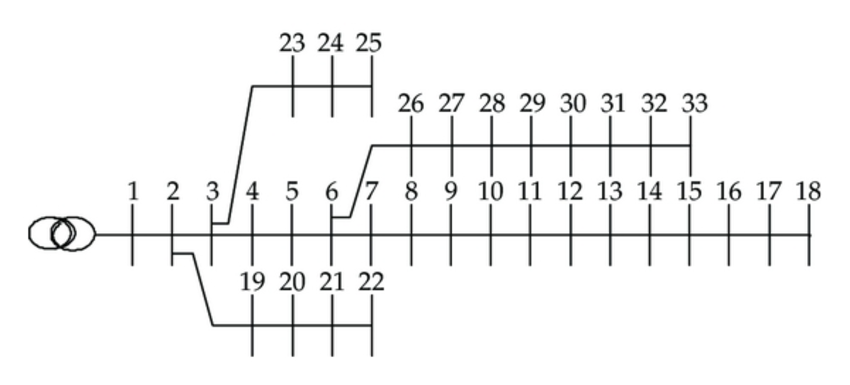}
\squeezeup
\caption{IEEE 33-bus system, where the loads at buses $\# 17, 20, 23, 26$ are replaced with the MDP given in Fig.~\ref{fig:TCL_MDP}. For more details about the MDP parameters, see Section~\ref{sec:data}. \label{fig:case33bw}}
\end{figure}

We test the proposed distribution power flow optimization with MDP and the proposed ST-D2 algorithm and its hybrid version on the IEEE 33-bus distribution system (Matpower `case33bw' case) \cite{89BWc,matpower}, see Fig.~\ref{fig:case33bw}.

\begin{figure}[t]
\centering
\hspace*{\fill}
\subfigure[]{\includegraphics[width=0.2\textwidth]{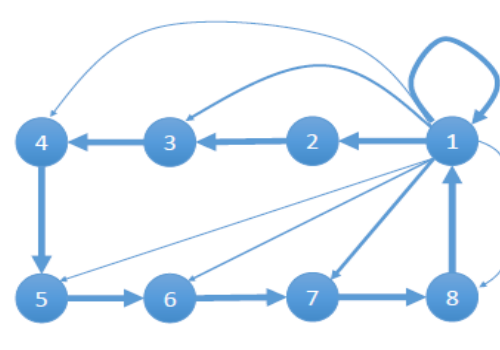}}\hspace*{\fill}
\subfigure[]{\includegraphics[width=0.29\textwidth, height= .2\textwidth]{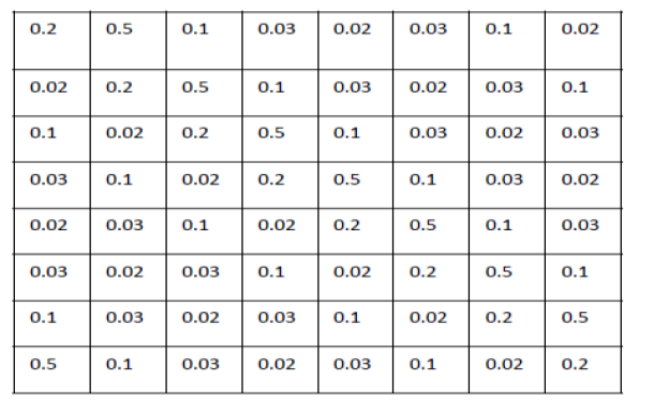}}
\squeezeup
\hspace*{\fill}
\caption{Parameters of the ``normal'' (e.g. user-defined) transition probability matrix $\overline{\mathcal{P}}$ for each MDP with $8$ states.
\label{fig:TCL_MDP}}
\end{figure}

We replace the loads at buses $\# 17, 20, 23, 26$ with TCL ensembles and model each ensemble using the MDP. Each TCL ensemble has $8$ states and the same target transition probability, $\overline{\mathcal{P}}$, shown in Fig.~\ref{fig:TCL_MDP}. Active and reactive loads associated with each state are given values from uniform intervals within $10\% - 200\%$ of the rated load at that node. \textcolor{black}{We consider the optimization horizon with 20 sequential  hourly intervals and} conduct experiments with a deterministic price computed as a constant (unity per MWh) with/without an additional random, time-dependent component, i.e. $u_t\sim 1+\mbox{rand}(t)$. Further, we consider two sets of penalty functions: (i) $\gamma^{\alpha\beta}=1$ for all transitions, as illustrated in Fig.~\ref{fig:TCL_MDP}(a), and (ii) $\gamma^{\alpha\beta}=1$ for one transition per state and $\gamma^{\alpha\beta}=10$ for all other transitions from that state. Using non-uniform penalty factors makes it possible to analyze the sensitivity of the MDP solution with respect to different transitions. In all simulations the optimal solution of each MDP problem is obtained using the forward-backward algorithm described in Appendix \ref{app:backforward}.
    \textcolor{black}{All instances presented below were solved using the Julia Jump optimization package \cite{julia_jump} with Gurobi solver v7.5 on Intel Core i5 1.6 GHz processor with 4 GB of RAM. All MPD problems were solved within 1 second, while the proposed algorithms converged within 1 minute. }

\subsection{MDP Optimization} 
\begin{figure}[bt]
\centering
%\hspace*{\fill}
\subfigure[]{\includegraphics[width=0.40\textwidth,height = 0.26\textwidth]{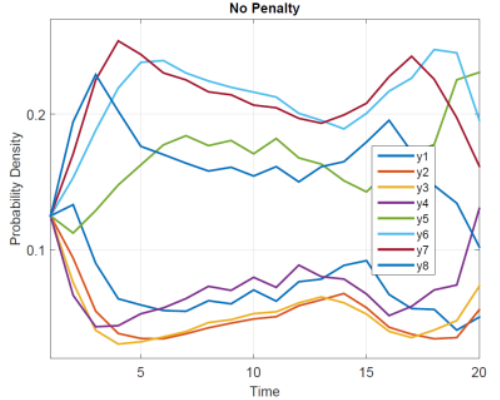}\label{fig:no_penalty}}\\%\hspace*{\fill}
\squeezeup
\subfigure[]{\includegraphics[width=0.40\textwidth,height = 0.26\textwidth]{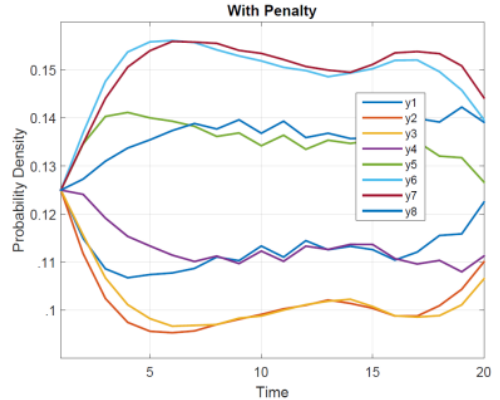}\label{fig:with_penalty}}
%\hspace*{\fill}
\squeezeup
\caption{Optimal solution $\rho$ of an individual MDP problem, i.e. for one TCL ensemble, with $8$ states over $20$ time steps with non-uniform costs and no network constraints (a) uniform $\gamma = 1$ (b) non-uniform $\gamma$ (see the description in Section~\ref{sec:data}).}
\end{figure}

Fig. \ref{fig:no_penalty} and \ref{fig:with_penalty} illustrate the effects of the penalty function $\gamma^{\alpha\beta}$ on the optimal operation of the individual MDP problem from Eq.~(\ref{hybrid_21a})-(\ref{hybrid_3a}) with non-uniform prices and without distribution power flow optimization. Each curve in these figures represents the optimal values of $\rho^\alpha(t)$ for different $\alpha$'s and $t$'s. In case of non-uniform values of penalty function $\gamma$, the optimized steady state probabilities differ significantly from the case with uniform values. First, imposing the non-uniform penalty function reduces the spread between the minimum and maximum values. Second, it smooths out the saw-tooth points, thus enabling a more fine-grained transition from one time instant to another.

\subsection{Distribution Power Flow Optimization with MDP} 
\begin{figure}[t]
\centering
\hspace*{\fill}
\includegraphics[width=0.52\textwidth,height = 0.3\textwidth]{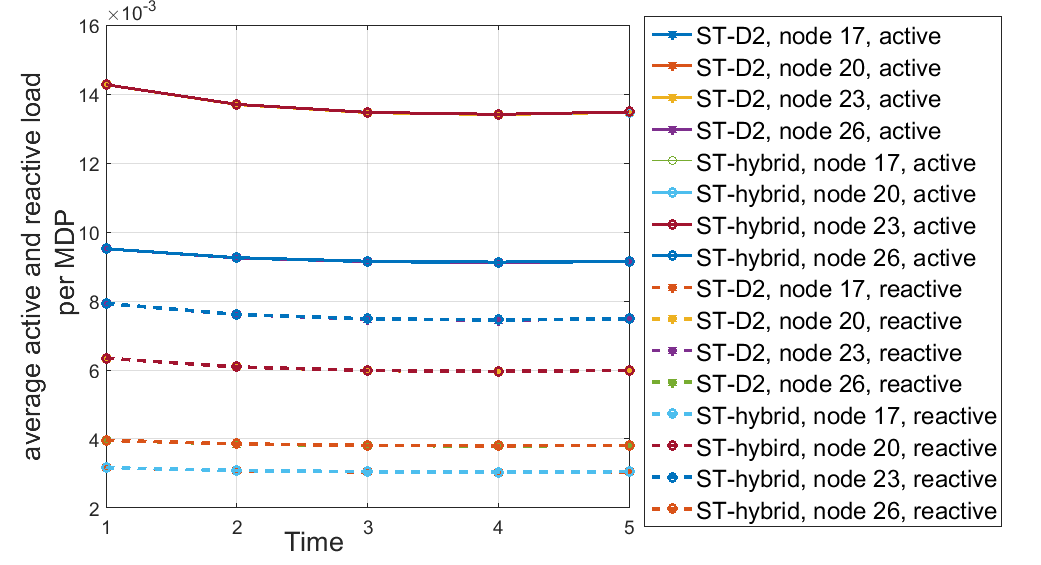}
\hspace*{\fill}
\squeezeup
\caption{The optimal MDP active and reactive power dispatch decisions on the TCL ensembles at different buses obtained with the ST-D2 and ST-Hybrid algorithms and non-uniform penalty function $\gamma$. These simulations include the distribution power flow optimization as given by Eq.~(6)-(15) over $5$ time intervals. Note that both algorithms yield identical decisions.}
\label{fig:loadsa}
\end{figure}

Next, we consider how the distribution power flow optimization affects the MDP control. To this end, we solve the integrated MDP and distribution power flow optimization in \eqref{hybrid_1}-\eqref{q_c} using the ST-D2 and ST-Hybrid algorithms described in Section \ref{sec:algo1}-\ref{sec:hybrid}. In our numerical experiments we did not observe any noticeable difference between the solutions of the two algorithms. As can be seen in Fig.~\ref{fig:loadsa} both algorithms return identical active and reactive power dispatch decisions on TCL ensembles located at different buses. Note that the dispatch decisions are identical in terms of the total power consumed and its distribution across the time intervals considered.

\begin{figure}[bt]
\centering
\subfigure[]{\includegraphics[width=0.40\textwidth,height = 0.28\textwidth]{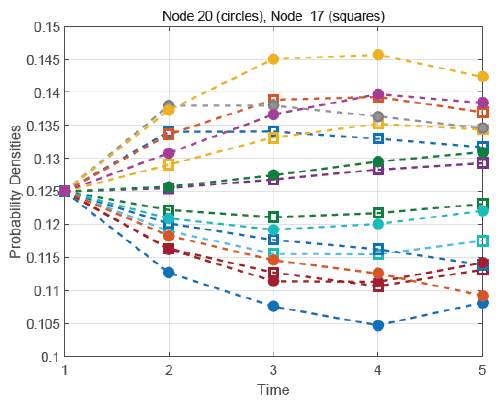}\label{fig:nonuniformMDP}}\hspace*{\fill}\\
\hspace*{\fill}
\subfigure[]{\includegraphics[width=0.40\textwidth,height = 0.26\textwidth]{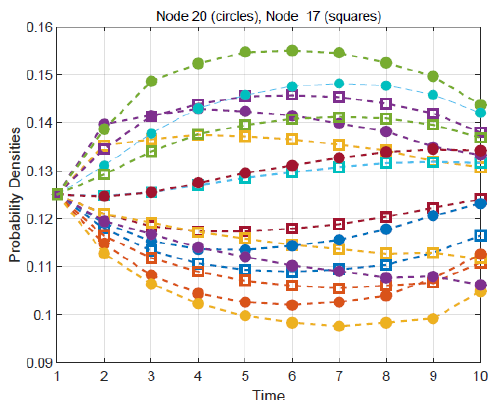}\label{fig:uniformMDP}}
\squeezeup
\hspace*{\fill}
%\vspace{-2pt}
\caption{Optimal solution $\rho$ for MDPs at nodes $17$ and $20$ found by the ST-Hybrid algorithm with non-uniform penalty $\gamma$ and network constraints and loss minimization objective according to Eqs. (5-14): (a) over $5$ time intervals with varying in time electricity price; (b) over $10$ time intervals with uniform electricity price.}
\end{figure}

To make our analysis more illustrative, the following discussion studies the optimized state probabilities on buses $17$ and $20$, where non-uniform and uniform values of the penalty function are enforced, respectively. Fig.~\ref{fig:nonuniformMDP} and \ref{fig:uniformMDP} display the optimized MDP decisions made at each node in case of uniform and non-uniform values of the penalty function. The optimized decisions vary between the two cases and as compared to Fig.~\ref{fig:no_penalty} and \ref{fig:with_penalty}. We attribute these changes to the following three effects. First, by considering line losses in power flows, the MDP decisions change to ensure loss minimization. Second, the MDP decisions are now constrained by the need to maintain the voltage and power flow limits. Third, the need to meet constraints on multiple TCL ensembles imposes restrictions on how their cumulative capacity can be used. These restrictions are implicitly accounted for in our optimization and allow for the optimal dispatch of the TCL ensembles.

%{\color{red} ... Deep, Yura ... it would be good to add some extra discussion of the results here ... or even better some extra studies ... for example we may want to find a regime where ignoring power flow part leads to dramatic violations of the voltage constraints and/or leads to very significant losses ... and other way around ... using not optimized transition probabilities results in a costly solution ... }

\section{Conclusion}

\label{sec:conclusion}

This paper presents a new approach to control multiple TCL ensembles located in the distribution system and to integrate these TCL ensembles with the distribution power flow optimization. This integration makes it possible to co-optimize the dispatch decisions on the TCL ensembles with the rest of the distribution operations, thus assisting the SO in reducing power losses and maintaining limits on power flows and nodal voltages. To solve the resulting optimization problem, we exploit spatial and temporal separability of TCL and distribution power optimization decisions and propose two decomposition algorithms. The proposed approach is tested on the IEEE 33-bus distribution system. We demonstrate that the optimal dispatch decisions on TCL ensembles are sensitive to the assumptions made on their comfort (utility) function and to the network constraints.
%\vspace{-10pt}

\appendices
\section{Backward-Forward Algorithm}
%\vspace{-5pt}
\label{app:backforward}
 The solution for MDP problem described in (\ref{hybrid_2a}) and Eq.~(\ref{profit_vs_welfare}) in general is described in detail in Appendix 1.9 in \cite{17CCb}. We present a brief overview of the \emph{ backward-forward} algorithm steps here:
\begin{itemize}
\item \underline{Backward in time step.} Starting with the final time, we solve for $\mathcal{P}$ recursively backward in time. In other words, at node $i$, optimal $\mathcal{P}^{\alpha\beta}_i(t)$'s for transitions to all states $\alpha$ from state $\beta$ at time $t$ is determined using optimal transitions in future times and associated cost functions. This is done either by a Lagrange relaxation or by minimizing a convex function.
\item \underline{Forward in time step.} We reconstruct $\rho$ running Eq.~(\ref{master-eq}) forward in time with the initial condition Eq.~(\ref{rho_init}).
\end{itemize}

Note that convexity of KL cost ensures that at each time in the backward step, a convex problem is solved with linear constraints due to stochasticity of $\mathcal{P}$ matrix.
%\begin{comment}
\section{On Probabilistic Description of the Site-Ensemble}
%\vspace{-5pt}
\label{app:large_n}
\textcolor{black}{For ensemble $i$ at the moment of time $t$ described by $\rho_i^\alpha(t)$, we generate $n_i$ i.i.d. samples, $\alpha_1,\cdots,\alpha_{n_i}$, whose total apparent power consumption is given by $\varsigma_{i}(t)=n_i^{-1}\sum_{k=1}^{n_i} \alpha_k$.} According to the law of large numbers, statistics of the apparent power consumption, $\varsigma_i(t)$, is Gaussian at $n_i\to\infty$ described by the following mean and variance
\begin{eqnarray}
&& \hspace{-0.5cm}\mathbb{E}\left[\varsigma\right]\to \sum_\alpha s_{i}^\alpha \rho^\alpha_i(t),\nonumber\\
&& \hspace{-0.5cm}\mbox{Var}\left[\varsigma\right]=\mathbb{E}\left[\left(\varsigma - \mathbb{E}\left[\varsigma\right]\right)^2\right]\to \sum_\alpha(s_{i}^\alpha-\sum_{\beta} s_{i}^\beta \rho^\beta_{i}(t))^2
\frac{\rho^\alpha_i(t)}{ n_i}
\nonumber%\label{variance_sigma}
\end{eqnarray}
%\end{comment}
%\vspace{-15pt}
\bibliographystyle{IEEEtran}
\bibliography{TCL,distcontrol}

\end{document}